\documentclass[12pt]{article}
\usepackage{graphicx}
\usepackage{epstopdf}
\usepackage{amsmath, amssymb}
\usepackage{shuffle}
\usepackage{color}
\usepackage{hyperref}
\hypersetup{bookmarksopen,bookmarksnumbered,pdfstartview=FitH,colorlinks,citecolor=blue,unicode}
\usepackage[numbers,sort&compress]{natbib}

\newcommand{\be}{\begin{equation}}
\newcommand{\ee}{\end{equation}}
\newcommand{\bea}{\begin{eqnarray}}
\newcommand{\eea}{\end{eqnarray}} 
\newcommand{\ba}{\begin{array}}
\newcommand{\ea}{\end{array}}

\newcommand{\half}{\frac{1}{2}}
\newcommand{\bb}{\bibitem}
\newcommand{\slsh}[1]{\displaystyle{\not} #1}

\begin{document}

\title{{\bf\color{blue} Chiral Superstring and CHY Amplitude}\\[-1.2in]
{\normalsize February 23, 2017\hfill YITP-SB-17-10}\\[1in]}
\date{}

\author{Yuqi Li$^{a}$\footnote{\href{mailto:yuqi.li@stonybrook.edu}{yuqi.li@stonybrook.edu}}\, , Warren Siegel$^{b}$\footnote{\href{mailto:siegel@insti.physics.sunysb.edu}{siegel@insti.physics.sunysb.edu}\vskip 0pt \hskip 8pt \href{http://insti.physics.sunysb.edu/\~siegel/plan.html}{http://insti.physics.sunysb.edu/$\sim$siegel/plan.html}} \bigskip\\ $^a$  \emph{Department of Physics and Astronomy}\\ \emph{State University of New York, Stony Brook, NY 11794-3840}\\
\\$^b$ \emph{C.~N.~Yang Institute for Theoretical Physics}\\ \emph{State University of New York, Stony Brook, NY 11794-3840}}
\maketitle

\begin{abstract}
\normalsize
We calculate the chiral string amplitude in pure spinor formalism and take four point amplitude as an example. The method could be easily generalized to $N$ point amplitude by complicated calculations. By doing the usual calculations of string theory first and using a special singular gauge limit, we produce the amplitude with the integral over Dirac $\delta$-functions. The Bosonic part of the amplitude matches the CHY amplitude and the Fermionic part gives us the supersymmetric generalization of CHY amplitude. Finally, we also check the dependence on boundary condition for heterotic chiral string amplitudes.

\end{abstract}

\newpage

\section{Introduction}
A few years earlier, Cachazo, He and Yuan (CHY) \cite{CH1,CH2,CH3,CH4} introduced one kind of string-like prescription of tree-level amplitudes of $N$ massless particles, in which they introduced one integral over only the $z$-dependent part and fixed the locations of each external line corresponding to $N$ particles by solutions of a series of formulas inside the $\delta$-functions. Those $N-3$ equations inside the $\delta$-functions are called ``Scattering Equations" by CHY, which have $(N-3)!$ solutions.

In this prescription, CHY formula looks very similar to some special string amplitudes with insertion of Dirac $\delta$-functions (see, e.g., \cite{DG,BDTV}). Shortly after, Mason and Skinner (MS) \cite{MS} introduced one kind of string amplitude with a $\delta$-function inserted into the vertex operators of their ambitwistor string theory. By taking the infinite-tension limit ($\alpha ' \rightarrow 0$), Berkovits \cite{B} almost immediately generalized the ambitwistor version to the pure spinor superstring version of scattering amplitude, and the Dirac $\delta$-functions were still inserted in the vertex operators.

Soon after, Siegel \cite{S1} introduced an approach closely related to the standard string theory by treating the singular worldsheet gauge (HSZ gauge \cite{HSZ}) as a singular limit with some simple modifications of the boundary conditions. Therefore, the integration over $\bar{z}$ with standard vertex operators under the HSZ gauge limit produces the CHY $\delta$-functions; that is, if we take the rules that first evaluate the amplitude in the usual conformal gauge and substitute the propagators with singular HSZ gauge limit before integrating over $\bar{z}$, the results after integration over $\bar{z}$ only depend on the $z$ coordinate with the correct number, namely $N-3$, of Dirac $\delta$-functions inserted in the integrand leaving the boundary conditions still simply modified. Furthermore, if we keep the modified boundary conditions and do the calculations in the usual conformal gauge, the scattering-equation $\delta$-functions are not seen explicitly, and the results are the same since they should be gauge independent \cite{S2}.

Recently, we are motivated by the fact that $\bar\delta(k\cdot P)$ insertion of MS ambitwistor string prescription could be treated as the (generalized) picture-changing operator;\footnote{Unlike usual CFT in string theory, we start with unintegrated version of vertex operator with ghost inserted, and then get the integrated $N-3$ vertex operators. So the picture-changing operators here are generalized picture-changing operators or ``inverse" picture-changing operators.} then, after BRST transformation, the calculation could be changed into formulas with the usual $\bar z$ dependence in $X$. Noticed that we use the integrated version in our calculations (such as $\int P^2$, $c_0=\int c$ and $b_0=\int b$) instead of the unintegrated version,
the ambitwistor string method \cite{ACS} would be BRST equivalent to the calculations in the normal string theory with some additional gauge limits, and the additional gauge limits lead to the localization of the external lines on the Riemann surface we integrated on by $\delta$-functions. The similar discussion could also be found in \cite{R} and our upcoming paper of loop-level calculations.

In this paper, we first discuss the gauge dependence of different prescriptions and calculate the four-point amplitude of massless states using the pure spinor formalism of Heterotic Strings with the singular HSZ gauge. Namely, we do the usual operator product expansions (OPEs) of four massless states under the pure spinor formalism and then substitute the singular limit of $\bar{z}$. The $\bar{z}$ integration of the Koba-Nielsen factor produces the CHY $\delta$-function and the current algebra part produces the Parke-Taylor-like factor by directly taking the singular gauge limit. The four-Boson part of the amplitude matches the usual Yang-Mills CHY formula and the two-Boson-two-Fermion and four-Fermion part of the amplitude naturally produce the supersymmetric version of CHY amplitude, namely super-Yang-Mills (SYM) CHY amplitude. In closed string calculations, by using the singular HSZ gauge limit before integration over $\bar{z}$, the CHY amplitude of four gravitons are produced with also the similar supersymmetric generalization as in the SYM case. Furthermore, the flip-sign method of the metric introduced by Huang, Siegel and Yuan (HSY) \cite{S2} could be checked in our calculations. Finally, the so-called bi-adjoint scalar amplitude of CHY formula could be simply produced by calculations of current algebras of closed strings.

\section{Gauge dependence}
The gauge dependence of chiral string theory here is of great importance. We first introduce $\mathcal{S}$ and $\mathcal{T}$ for short:
\begin{eqnarray}
\mathcal{S}&=&\{Q,(b-\bar b)\}=X'\cdot P\nonumber\\
\mathcal{T}&=&\{Q,(b+\bar b)\}=\frac{1}{2}(P^2+\frac{1}{\alpha'^2}X'^2)
\end{eqnarray}
In the usual string field theory, $\delta(f)/\{Q,f\}$ and $\delta(f)\delta\{Q,f\}$ could be expressed as (see also \cite{Berk, Marnelius}):
\begin{eqnarray}
\frac{\delta(f)}{\{Q,f\}}&=&\int_0^\infty d\tau\, d\tilde{c}\, e^{\tilde{c}f-\tau\{Q,f\}}\nonumber\\
\delta(f)\delta(\{Q,f\})&=&\oint_{0}^{2\pi} \frac{d\sigma}{2\pi i}\,d\tilde{c}\,e^{\tilde{c}f-\sigma\{Q,f\}}
\end{eqnarray}
Identify $f$ with either $b+\bar b$ or $b-\bar b$ and use $\delta(b\pm\bar b)=b\pm \bar b$. Thus, like the usual "plumbing" analogy in string field theory to get CFT (or, if you like, think of changing from interaction picture to Heisenberg picture), we have $N$ unintegrated vertex operators with $c$ ghost insertions in vertex $V$s,
\begin{eqnarray}
U=(c+\bar c)(c-\bar c)V
\end{eqnarray}
Here, we consider $V\sim e^{ik\cdot X}$ in this section. The propagator with respect to $\mathcal{S}$ and $\mathcal{T}$ is also easy to calculate:
\begin{eqnarray}
\Delta=\frac{b_0+\bar b_0}{\mathcal{T}_0}(b_0-\bar b_0)\delta(\mathcal{S}_0)
\end{eqnarray}
Thus, one would get the integrated vertex operators by sandwiching two propagators to cancel the $c$ ghosts and leave the integrations over the corresponding space, in this case $(\sigma,\tau)$ space.
\subsection{Conformal gauge}
To change into chiral string boundary conditions in conformal gauge, one would effectively get the sign-change of the $b$ ghosts, namely,
$$\bar b\rightarrow-\bar b$$
which is also effectively equivalent to switching the role of $\mathcal{S}$ and $\mathcal{T}$.
Therefore, the propagator changes into
\begin{eqnarray}
\Delta_c=\frac{b_0-\bar b_0}{\mathcal{S}_0}(b_0+\bar b_0)\delta(\mathcal{T}_0)
\end{eqnarray}
while the unintegrated vertex operators $U$ are the same.
Here, one still has the fact that the propagators acting on unintegrated vertex operators gives integrated vertex operators.

\subsection{HSZ gauge}
HSZ gauge is a singular gauge choice which is not conformal as the gauge choice discussed above.
We first take the conformal gauge of chiral strings with the same propagators $\Delta_c$ and then use the HSZ gauge before integration. As shown in details in later discussions of this paper, the HSZ gauge is to modify the propagator $\langle X X\rangle$ up to an appropriate regulator by simply transforming the coordinates:
\begin{eqnarray}
z&\rightarrow& z\nonumber\\
\bar z&\rightarrow& \bar z-\beta z
\end{eqnarray}
The CHY $\delta$-function would appear after integration over $\bar z$.

Noted that we still have $\mathcal{S}$ and $\mathcal{T}$ playing the roles as propagators. The additional singular gauge choices provide the localization of positions of each vertex operator and the integrations over $\bar z$ express the localization in terms of $\delta$-functions.

\subsection{MS gauge}
By choosing the same gauge of $\mathcal{S}$ (Siegel gauge) as usual but different gauge choice of $\mathcal{T}$, MS formalism has the propagator only depending on $\mathcal{S}$.
\begin{eqnarray}
\Delta_{MS}=\frac{b_0-\bar b_0}{\mathcal{S}_0}
\end{eqnarray}
Since the unintegrated vertex operators are sandwiched by propagators $\Delta_{MS}$ to get integrated vertex operators, one only obtains the $\sigma$ integrations (after relabeling, $\sigma$ is changed to be $z$) and leaves the $\mathcal{T}$ separated into vertex operators insertions, which is effectively the inverse picture-changing operator:
\begin{eqnarray}
\Upsilon=(b+\bar b)\delta(\mathcal{T})
\end{eqnarray}
The propagator $\Delta_{MS}$ and inverse picture-changing operator $\Upsilon$ together cancel the $c$ ghosts of the unintegrated vertex operators ($U_O$ and $\bar U_O$ correspond to open string):
$$U=(c+\bar c)(c-\bar c)e^{ik\cdot X}=(c+\bar c)(c-\bar c) V=U_O \bar{U}_O.$$ 
Before sandwiching the propagators $\Delta_{MS}$ on both sides, the inverse picture-changing operator collides on the unintegrated vertex operator first:
\begin{eqnarray}\label{PCO1}
\mathcal{W}=\lim_{\epsilon\rightarrow0}\Upsilon(\epsilon) U(0)
\end{eqnarray}
Using $\mathcal{S}$ as the regulator to regularize the separation of $\epsilon$, one gains \footnote{Since $\mathcal{S}$ provides all the propagations, $X(\sigma_1)\cdot P(\sigma_2)\sim\frac{1}{\sigma_{1}-\sigma_2}$, $b(\sigma_1)c(\sigma_2)\sim \frac{1}{\sigma_{1}-\sigma_2}$ and $\bar b(\sigma_1)\bar c(\sigma_2)\sim \frac{1}{\sigma_{1}-\sigma_2}$ are used and $\sigma$ is relabeled as $z$ in this subsection.}
\begin{eqnarray}\label{delta}
&&\lim_{\epsilon\rightarrow0}\delta(\mathcal{T} (\epsilon))e^{ik\cdot X(0)}\nonumber\\
&=&\lim_{\epsilon\rightarrow0}\int d\tau\, e^{\tau \mathcal{T} (\epsilon)} \, e^{ik\cdot X(0)}\nonumber\\
&=&\lim_{\epsilon\rightarrow0}\int d\tau :e^{\tau \mathcal{T} (\epsilon)+ik\cdot X(0)+\frac{\tau k\cdot P}{\epsilon}}:\nonumber\\
&=&\lim_{\epsilon\rightarrow0}\,(\delta(k\cdot P(\epsilon))\epsilon+\mathcal{O}(\epsilon^2)) \,e^{ik\cdot X(0)}\nonumber
\end{eqnarray}
Then, use
\begin{eqnarray}\label{1}
(b +\bar b )(\sigma_1)(c +\bar c )(\sigma_2)\sim\frac{1}{\sigma_1-\sigma_2}= \frac{1}{\epsilon}
\end{eqnarray}
to get
\begin{eqnarray}
\mathcal{W}&=&\lim_{\epsilon\rightarrow0}\Upsilon(\epsilon) U(0)=\lim_{\epsilon\rightarrow0}\, \frac{1}{\epsilon}(\delta(k\cdot P(\epsilon))\epsilon+\mathcal{O}(\epsilon^2))(c-\bar c) \, e^{ik\cdot X(0)}\nonumber\\
&\sim& (c-\bar c)\delta(k\cdot P(0)) \, e^{ik\cdot X(0)}
\end{eqnarray}
Then, sandwiching those vertex operators by the propagators, 
\begin{eqnarray}\label{sandwich}
\Delta_{MS} \mathcal{W}\Delta_{MS}\mathcal{W}\Delta_{MS}\mathcal{W}\Delta_{MS}\dots,
\end{eqnarray} 
would lead to the MS formalism:\\
(a) $(b_0-\bar b_0)$ cancels $(c-\bar c)$ in $\mathcal{W}$; namely, for the ghost part
\begin{eqnarray}\label{sandwich2}
\{\mathcal{W},b_0-\bar b_0\}&=&V\nonumber\\
\Rightarrow(b_0-\bar b_0)\mathcal{W}(b_0-\bar b_0)&=&V(b_0-\bar b_0)
\end{eqnarray}
(b) $\mathcal{S}_0$ provides the $z$-integrations at each point $z$ of integrated vertex operator after relabeling $\sigma$ to be $z$.\\
Noted that there is no need of $\alpha '$ limit in this calculation.
\subsection{MS gauge in pure spinor formalism}
When calculate the MS prescription in pure spinor formalism, the $b$ ghost in pure spinor formalism is composite (there is no $c$ ghost in pure spinor formalism).

For open string, considering
\begin{eqnarray*}
\{Q,b(z_1)U(z_2)\}&=&T(z_1)U(z_2)\nonumber\\
&\sim&\frac{1}{z_1-z_2}\partial U(z_2)\nonumber\\
&=&\frac{1}{z_1-z_2}[Q,V(z_2)],
\end{eqnarray*}
We already used $[Q,V]=\partial U$ in the last line of calculation. One gets (the BRST trivial part is absorbed into the commutator with $Q$)
\begin{eqnarray}\label{PCO}
b(z_1)U(z_2)\approx\frac{1}{z_1-z_2}V(z_2)+[Q,\dots]
\end{eqnarray}
or more general
\begin{eqnarray}\label{PCO_}
(b(z_1)+w(z_1))(Q+U(z_2))\approx T(z_1)+\frac{1}{z_1-z_2}V(z_2).
\end{eqnarray}
The BRST trivial term in \eqref{PCO} is due to the background contribution to $b$ in \eqref{PCO_}; furthermore, $w$, $U$ and $V$ are the background contributions to $b$, $Q$ and $T$, respectively.

Noticed that the former calculations also imply
\begin{eqnarray*}
\{U,b_0\}=V
\end{eqnarray*}
and further
\begin{eqnarray}
b_0 U b_0=Vb_0.
\end{eqnarray}

For closed strings, vertex operators and ghosts could be separated into two parts, which gives:
\begin{eqnarray}
&&(b+\bar b)U\bar U\sim\frac{1}{z}(V\bar U-U\bar V)\\
&&\{b_0-\bar b_0,V\bar U-U\bar V\}\sim V \bar V
\end{eqnarray}
for vertex insertions (like \eqref{1} in Bosonic case) and propagators (analogous to \eqref{sandwich2}), respectively. And the discussion of $\delta(\mathcal{T})$ insertion of the integrated vertex operators is the same.


\section{Review of pure spinor formalism}
\subsection{Conformal field theory}
Pure spinor formalism is based on a worldsheet conformal field theory (CFT) with fields $X^{m}, \theta^{\alpha}$ and ghost $\lambda^{\alpha}$ with the corresponding conjugate momenta, $m=0,\dots,9$ and $\alpha=1,\dots,16$ as in the usual pure spinor formalism. Then, the worldsheet action in a flat background is given by \cite{B, Berk2, Berk3}:
\begin{equation}
S=\int d^2 z\, (\half \partial X^m \bar\partial X_m + p_\alpha \bar\partial\theta^\alpha 
+ \bar b \partial\bar c) + S_\lambda+ S_J, 
\end{equation}
where $S_\lambda$ and $S_J$ are the actions for $\lambda^\alpha$ and $J^I$.
And the ghost is constraint further by
\begin{align}
\lambda\gamma^{m} \lambda= 0 \,.  
\end{align}
With the definition
\begin{align} 
d_\alpha&=p_\alpha-\half (\gamma^m\theta)_\alpha \partial X_m -\frac{1}{8} (\gamma^m\theta)_\alpha
(\theta\gamma_m\partial\theta)\\ 
\Pi^m&=\partial X^m+\half(\theta\gamma^m\partial\theta)~.
\end{align}
one can easily get the OPEs of those fields:
\begin{align}
X^m(y)X^n(z)&\sim  -\half\alpha' \eta^{mn}ln|y-z|^2 \, , & 
p_\alpha(y)\theta^\beta(z)&\sim\frac{\alpha'\delta_\alpha^\beta}{y-z}\\
d_\alpha(y) d_\beta(z)&\sim -\frac{\alpha'}{y-z}\gamma^m_{\alpha\beta}\Pi_m  \,, &
d_\alpha (y)\Pi^m(z)&\sim \frac{\alpha'}{y-z}(\gamma^m\partial\theta)_\alpha\\
d_\alpha(y)\partial\theta^\beta(z)&\sim \frac{\alpha'}{(y-z)^2}\delta_\alpha^\beta  \,, &
\Pi^m(y)\Pi^n(z)&\sim -\frac{\alpha'}{(y-z)^2}\eta^{mn}~.
\end{align}
and the OPEs of current algebra correlated to $S_J$:
\be
J^I(y)J^K(z)\sim \frac{k\delta^{IK}}{(y-z)^2}+f_L^{IK}\frac{J^L(z)}{(y-z)}
\ee
here $f_L^{IK}$ is the gauge group structure constant. For any superfield $\mathcal{F}(X^m(z),\theta^\alpha(z))$, the OPEs satisfy as follows:
\begin{align}
\Pi^m(y)\mathcal{F}(z)&\sim - \frac{\alpha'}{y-z}\partial^m \mathcal{F}(z) \,\\
d_\alpha(y) \mathcal{F}(z)&\sim \frac{\alpha'}{y-z}D_\alpha \mathcal{F}(z)  ~
\end{align}
with $D_\alpha:= \frac{\partial}{\partial\theta^\alpha}+\half (\gamma^m\theta)_\alpha \partial_m$ the superderivative. And recall:
\begin{eqnarray}
A_\alpha&=& \frac{1}{2} (\theta \gamma^m)_\alpha \epsilon_m +\frac{1}{3}
(\theta \gamma^m)_\alpha (\theta \gamma_m \xi)+\frac{1}{16}(\theta \gamma^m)_\alpha (\theta
\gamma_m{}^{pq} \theta)\partial_q \epsilon_p + \dots \nonumber\\
A_m&=&\epsilon_m+(\theta \gamma_m \xi)+\frac{1}{4} (\theta
\gamma_m^{pq} \theta) \partial_q \epsilon_p+\frac{1}{12} (\theta
\gamma_m^{qp} \theta) (\theta \gamma_q\partial_p \xi)+\dots\nonumber\\
W^\alpha&=&\frac{1}{10}\gamma_m^{\alpha\beta}(D_\beta A^m-\partial^mA_\beta)\nonumber\\
F_{mn}&=&\partial_{[m}A_{n]}
\end{eqnarray}
Here, $\epsilon_m$ is the gluon polarization vector and $\xi_\alpha$ is the wavefunction for gluino.
Thus, the vertex operators of open strings, closed strings and heterotic strings could be easily written as:\\
Open:
\begin{align}
U_O&=e^{ik\cdot X(z)}\lambda^\alpha A_\alpha(\theta) \nonumber\\
V_O&=e^{ik\cdot X(z)}(\partial\theta^\alpha A_\alpha +\Pi^m A_m +d_\alpha W^\alpha +\half N^{mn} F_{mn} )
\end{align}
Closed:
\begin{align}
U_{C}&:=e^{ik\cdot X(z,\bar z)} \lambda^\alpha A_\alpha(\theta) \lambda^{\bar\beta} A_{\bar\beta}(\bar \theta)   \nonumber\\
V_{C}&:=e^{ik\cdot X(z,\bar z)} (\partial\theta^\alpha A_\alpha +\Pi^m A_m +d_\alpha W^\alpha
+\half N^{mn} F_{mn}) \nonumber\\ 
&~~\otimes(\bar \partial\theta^{\bar\beta} A_{\bar\beta} +\bar\Pi^m \bar A_m 
+d_{\bar\beta} W^{\bar\beta} +\half \bar N^{mn} \bar F_{mn})~.
\end{align}
Heterotic:
\begin{align}\label{heterotic}
U_H&:=e^{ik\cdot X(z,\bar z)}\lambda^\alpha A_{\alpha I}(X,\theta) \bar c \bar J^I \nonumber\\
V_H:&=e^{ik\cdot X(z,\bar z)}(\partial\theta^\alpha A_{\alpha I} +\Pi^m A_{m I} +d_\alpha W^\alpha_I +\half N^{mn} F_{mn I})\bar J^I
\end{align}
The heterotic string vertex operators \eqref{heterotic} are only for the super-Yang-Mills amplitude but not for supergravity amplitude. All $U$s stand for the unintegrated vertex operator and all $V$s are for integrated vertex operators. Here, $I$ indices are group index. Thus, the expressions of four point amplitude of closed (heterotic) strings are:\footnote{We fix three points for unintegrated vertices at $z_1$, $z_2$ and $z_3$.}
\begin{align}
\mathcal{A}_4=\langle U_{C(H)1}(z_1)U_{(C(H)2)}(z_2)U_{C(H)3}(z_3)\int d^2z_4\, V_{C(H)4}(z_4) \rangle ~.
\end{align}
We also need the useful notations in the following calculations:
\begin{align}
\mathcal{A}_{KN}(z_{ij})=\langle
e^{ik_1\cdot X(z_1,\bar z_1)}e^{ik_2\cdot X(z_2),\bar z_2}e^{ik_3\cdot X(z_3,\bar z_3)}
e^{ik_4\cdot X(z_4,\bar z_4)} \rangle  =
\prod_{i<j}^4 |z_i-z_j|^{\alpha' k_i\cdot k_j}.
\end{align}
and correlation functions of currents and ghosts
\begin{eqnarray}
\mathcal{A}_{J}&=&\langle  c J(z_1) c J(z_2)  c J( z_3) J( z_4) \rangle\nonumber\\
\mathcal{A}_{\bar J}&=&\langle  \bar c J(\bar z_1) \bar c J(\bar z_2)  \bar c J(\bar z_3) J(\bar z_4) \rangle.
\end{eqnarray}

\subsection{Heterotic superstring amplitude}
The four point heterotic string amplitude is given by:
\begin{align}
\mathcal{A}_{H4}=\langle U_{H1}(z_1)U_{H2}(z_2)U_{H3}(z_3)\int d^2z_4 V_{H4}(z_4) \rangle ~.
\end{align}
Using the OPEs listed above, the amplitude could be factorized into:

\begin{align}
\mathcal{A}_{H4}=\int d^2 z_4\,\, \mathcal{A}_{KN}(z_{ij})\mathcal{A}_J(\bar z_4)\mathcal{A}_O(z_4).
\end{align}
Here, $\mathcal{A}_O$, the open superstring correlation functions without the Koba-Nielsen factor, could be factorized further:

\begin{equation}
\mathcal{A}_O=\mathcal{A}_{4B}+\mathcal{A}_{2B2F}+\mathcal{A}_{4F},
\end{equation}
$\mathcal{A}_{4B},\mathcal{A}_{2B2F}$ and $\mathcal{A}_{4F}$ correspond to the four-Boson, two-Boson-two-Fermion and four-Fermion case, namely,
\begin{eqnarray}
\mathcal{A}_{4B}&=&\frac{\alpha'}{z_1-z_4}( k_{1} \cdot \epsilon_{4} \, k_{2} \cdot \epsilon_{3} \epsilon_{1} \cdot \epsilon_{2} 
+ k_{3}  \cdot \epsilon_{4} \, k_{2} \cdot \epsilon_{1} \, \epsilon_{2} \cdot \epsilon_{3} 
+ k_{1} \cdot \epsilon_{3} \, k_{3} \cdot \epsilon_{2} \, \epsilon_{1} \cdot \epsilon_{4}\nonumber\\
&&- k_{4} \cdot \epsilon_{1} \, k_{2} \cdot \epsilon_{3} \, \epsilon_{2} \cdot \epsilon_{4}   
+ k_{1} \cdot k_{4} \, \epsilon_{1} \cdot \epsilon_{2} \, \epsilon_{3} \cdot \epsilon_{4} 
- k_{2}\cdot k_{4} \, \epsilon_{2} \cdot \epsilon_{3} \, \epsilon_{1} \cdot \epsilon_{4})\nonumber \\
&&+cyclic
\end{eqnarray}
\begin{eqnarray}
\mathcal{A}_{2B2F}&=&\frac{\alpha'}{z_1-z_4}(i\xi_{1} \slsh{\epsilon_{2}} \xi_{4} \, k_{2} \cdot \epsilon_{3}  \, 
	+  i\xi_{2} \slsh{\epsilon_{3}} \xi_{4} \, k_{4} \cdot \epsilon_{1}  \,
	+  i \xi_{1} \slsh{k_{3}} \xi_{4}\epsilon_{2} \cdot \epsilon_{3}   
	+ i\xi_{3} \slsh{\epsilon_{2}} \slsh{\epsilon_{1}} \slsh{k_{1}} \xi_{4}\nonumber\\
	&&+i\xi_{2} \slsh{\epsilon_{1}} \xi_{3} \, k_{1} \cdot \epsilon_{4}  \, 
-i\xi_{2} \slsh{\epsilon_{4}} \xi_{3} \, k_{4} \cdot \epsilon_{1}  \, 
	+  i\xi_{1} \slsh{\epsilon_{3}} \xi_{2} \, k_{1} \cdot \epsilon_{4}  \,
	+   i\xi_{2} \slsh{k_{4}} \xi_{3}\epsilon_{1} \cdot \epsilon_{4} - i\xi_{3} \slsh{\epsilon_{2}} \slsh{\epsilon_{4}} \slsh{k_{4}} \xi_{1})  \nonumber\\
	&&+cyclic
\end{eqnarray}
and
\begin{eqnarray}
\mathcal{A}_{4F}=\frac{-\alpha'}{z_1-z_4} 
(\xi_1\gamma^m\xi_4)(\xi_2\gamma_m\xi_3)+cyclic
\end{eqnarray}
here we omit the overall constant factor $\frac{1}{5760}$ which will be irrelevant to our discussions.\footnote{Chan-Paton factors are trivial and we omit the discussion here.}

\subsection{Closed string amplitude}
For the closed string, we get the expressions of amplitude
\begin{align}
\mathcal{A}_{C4}=\langle U_{C1}(z_1)U_{C2}(z_2)U_{C3}(z_3)\int d^2z_4 \, V_{C4}(z_4) \rangle ~.
\end{align}
and it is easy to get the integral as:

\begin{align}
\mathcal{A}_{C4}=\int d^2 z_4\, \mathcal{A}_{KN}(z_{ij})(\mathcal{A}_O(z_4)\otimes\mathcal{A}_O(\bar z_4))
\end{align}
We could also factorize more
\begin{eqnarray}
\mathcal{A}_O(z_4)\otimes\mathcal{A}_O(\bar z_4)=\mathcal{A}_{4B}(z_4)\mathcal{A}_{4B}(\bar z_4)+\mathcal{A}_{2B2F}(z_4)\mathcal{A}_{2B2F}(\bar z_4)+\mathcal{A}_{4F}(z_4)\mathcal{A}_{4F}(\bar z_4)
\end{eqnarray}
to obtain
\begin{eqnarray}
\mathcal{A}_{C4}=\int d^2 z_4\, \mathcal{A}_{KN}(z_{ij})(\mathcal{A}_{4B}(z_4)\mathcal{A}_{4B}(\bar z_4)+\mathcal{A}_{2B2F}(z_4)\mathcal{A}_{2B2F}(\bar z_4)+\mathcal{A}_{4F}(z_4)\mathcal{A}_{4F}(\bar z_4))
\end{eqnarray}
Here, the expressions of $\mathcal{A}_{4B},\mathcal{A}_{2B2F}$ and $\mathcal{A}_{4F}$ are the same as those in the previous subsection, and the anti-holomorphic part is just the interchange $z \rightarrow \bar z$. The calculation agrees with \cite{LS}.

\section{HSZ gauge and CHY formula}

Before integrating over $\bar z$, we need to take the singular HSZ gauge limit. As shown in \cite{S1}, we take the HSZ gauge instead of the usual conformal gauge:
\begin{eqnarray}\label{HSZ}
z&\rightarrow& z\nonumber\\
\bar z&\rightarrow& \bar z-\beta z.
\end{eqnarray}
Then the modification of conformal gauge propagator $\langle XX\rangle$ leads to the substitution:
\begin{eqnarray}
\frac{1}{z}&\rightarrow&\frac{1}{z}\nonumber\\
\frac{1}{\bar z}&\rightarrow&\frac{1}{\beta z}+\frac{1}{\beta^2}\frac{\bar z}{z^2}.
\end{eqnarray}
When $\beta \rightarrow \infty$, we only keep the highest order of $\beta$, namely,
\begin{eqnarray}\label{limit}
\frac{1}{\bar z}&\rightarrow&\frac{1}{\beta z}.
\end{eqnarray}
Note that we would have a minus in front (up to a regulator), if we naively take the HSZ gauge limit \eqref{HSZ}:
\begin{eqnarray}
\frac{1}{\bar z}\rightarrow-(\frac{1}{\beta z}+\frac{1}{\beta^2}\frac{\bar z}{z^2})
\end{eqnarray}
but this minus sign is already absorbed into the ``sign-flip" caused by the change of boundary conditions as shown by HSY \cite{S2} (see also \eqref{sign}).

First, change the boundary condition, which receives a minus sign in front of the logarithm of $\bar z$ part,
\begin{eqnarray}
\ln|z_{ij}|&=&\half(\ln z_{ij}+\ln \bar z_{ij})\nonumber\\
&\rightarrow&\half(\ln z_{ij}-\ln \bar z_{ij})\nonumber\\
&=&\half\ln\frac{z_{ij}}{\bar z_{ij}},
\end{eqnarray}
and substitute the HSZ gauge into the Koba-Nielsen factor, 
\begin{eqnarray}
\ln\frac{z_{ij}}{\bar z_{ij}}&\rightarrow&\ln (1+\frac{\bar z_{ij}}{\beta z_{ij}})\nonumber\\
&\sim&\frac{\bar z_{ij}}{\beta \, z_{ij}},
\end{eqnarray}
then, we obtain the reduced form of Koba-Nielsen factor:
\begin{eqnarray}
\mathcal{A}_{KN}&=&\prod_{i<j}^4 |z_i-z_j|^{\alpha' k_i\cdot k_j}\nonumber\\
&=&\exp(\sum_{i<j}\alpha'k_i\cdot k_j\ln|z_i-z_j|)\nonumber\\
&\rightarrow&\exp(\sum_{i<j}\half\alpha'k_i\cdot k_j\ln\frac{z_{ij}}{\bar z_{ij}})\nonumber\\
&\sim&\exp(\sum_{i<j}\half\alpha'k_i\cdot k_j\frac{\bar z_{ij}}{\beta z_{ij}})\nonumber\\
&=&\exp(\frac{\alpha '}{\beta}\bar z_4\sum_{j=1}^3\frac{k_4\cdot k_j}{z_{4j}})
\end{eqnarray}
here, $z_{ij}=z_i-z_j$ and $\bar z_{ij}=\bar z_i-\bar z_j$. It is almost the CHY scattering equations in $N=4$ case.

Second, we do the same substitution for $\mathcal{A}_{O}(\bar z_4)$. Recall that
$$\frac{1}{\bar z}\rightarrow\frac{1}{\beta z}$$
we have also
\begin{eqnarray}
\frac{1}{\bar z_{ij}}&\rightarrow&\frac{1}{\beta z_{ij}}
\end{eqnarray}
This leads to the effect $\bar z\rightarrow z$ with an overall factor $\frac{1}{\beta}$ in front:
\begin{eqnarray}
\mathcal{A}_{O}(\bar z_4)=\frac{1}{\beta}\mathcal{A}_O(z_4)
\end{eqnarray}
Furthermore, the substitution for current part $\mathcal{A}_J(\bar z_4)$ gives us the similar expressions: \footnote{We omit the Chan-Paton factor, which is a single trace, here also.}
\begin{eqnarray}
\mathcal{A}_J(\bar z_4)&=&\bar z_{12}\bar z_{23}\bar z_{31}\frac{1}{\bar z_{12}\bar z_{23}\bar z_{34}\bar z_{41}}\nonumber\\
&\sim&\frac{1}{\beta}z_{12}z_{23}z_{31}\frac{1}{z_{12}z_{23}z_{34}z_{41}}\nonumber\\
&=&\frac{1}{\beta}\mathcal{A}_{J}(z_4)
\end{eqnarray}

Finally, integration over $\bar z$ will only affect the Koba-Nielsen factor and give us:

\begin{eqnarray}
\int d\bar z_4 \,\exp{(\frac{1}{\beta}\bar z_4\sum_{j=1}^3\frac{\alpha 'k_4\cdot k_j}{z_{4j}})}=\frac{\beta}{\alpha '}\,\,\delta(\sum_{j=1}^3\frac{k_4\cdot k_j}{z_{4j}})
\end{eqnarray}
which produces the CHY $\delta$-function.

In conclusion, the heterotic string amplitude is

\begin{eqnarray}
\mathcal{A}_{H4}&=&\int dz_4\, \frac{\beta}{\alpha'}\delta(\sum_{j=1}^3\frac{k_4\cdot k_j}{z_{4j}})\frac{1}{\beta}\mathcal{A}_J(z_4)\mathcal{A}_O(z_4)\nonumber\\
&=&\frac{1}{\alpha'}\int dz_4\, \delta(\sum_{j=1}^3\frac{k_4\cdot k_j}{z_{4j}})\mathcal{A}_J(z_4)\mathcal{A}_O(z_4)\nonumber\\
\end{eqnarray}
It is easily to check that four-Boson case matches with the usual CHY Yang-Mills amplitude. 

Furthermore, one could get the closed string amplitude straightforward:

\begin{eqnarray}
\mathcal{A}_{C4}&=&\int dz_4\, \frac{\beta}{\alpha'}\,\delta(\sum_{j=1}^3\frac{k_4\cdot k_j}{z_{4j}})\mathcal{A}_O(z_4)\frac{1}{\beta}\mathcal{A}_O(z_4)\nonumber\\
&=&\frac{1}{\alpha'}\int dz_4\, \delta(\sum_{j=1}^3\frac{k_4\cdot k_j}{z_{4j}})|\mathcal{A}_O(z_4)|^2
\end{eqnarray}
whose four-Boson part matches with CHY Graviton amplitude.

If we now introduce the following vertex operators,
\begin{eqnarray}
U_S&=&e^{ik\cdot X(z,\bar z)}cJ(z_4)\bar c\bar J(\bar z_4)\nonumber\\
V_S&=&e^{ik\cdot X(z,\bar z)}J(z_4)\bar J(\bar z_4),
\end{eqnarray}
the four point amplitude with respect to those vertex operators are obtained by simply calculations:

\begin{eqnarray}
\mathcal{A}_{S}&=&\langle U_{S1}(z_1)U_{S2}(z_2)U_{S3}(z_3)\int d^2z_4\, V_{S4}(z_4) \rangle \nonumber\\
&=&\int d^2z_4\,\mathcal{A}_J(z_4)\mathcal{A}(\bar z_4)\nonumber\\
&=&\frac{1}{\alpha'}\int dz_4\, \delta(\sum_{j=1}^3\frac{k_4\cdot k_j}{z_{4j}}) (\mathcal{A}_J(z_4))^2
\end{eqnarray}
which is the CHY scalar amplitude.

Note that the $\beta$ dependence is cancelled by the $\delta$ functions and OPEs; the amplitudes are gauge independent so as they should be. Since we already omit the coupling constants in our calculations, the residual $\alpha '$ dependence could be further absorbed into the definition of coupling constants, namely,
\begin{eqnarray}
g_s \phi^3 &\,& g_s\sim\frac{1}{\sqrt{\alpha'}}\nonumber\\
g_A A^2\partial A &\,& g_A\sim 1\nonumber\\
g_h h^2\partial^2 h &\,& g_h\sim\sqrt{\alpha'}
\end{eqnarray}
with respect to scalar, vector and graviton fields.
\section{Boundary conditions}
It is already shown by \cite{S1} and further \cite{S2} that the change of boundary conditions would leave a sign-flip of the metric. In \cite{S2}, one does not need to take the HSZ gauge limit, but here one needs to check whether the statements are still true when taking the HSZ gauge, which could be checked as follows:\\
(1) The sign-flip holds not only for the spacetime metric $\eta^{mn}$ but also for $\delta^\beta_\alpha$.\\
(2) Those sign-flips will affect the OPEs related to the integrated vertex operators.\\
(3) The net effect of HSZ gauge limit reads:
\begin{eqnarray}\label{sign}
\frac{\eta^{mn}}{\bar z_i-\bar z_j}&\rightarrow&\frac{(-)\eta^{mn}}{-\beta(z_i-z_j)}=\frac{\eta^{mn}}{\beta(z_i-z_j)}\nonumber\\
\frac{\delta^\beta_\alpha}{\bar z_i-\bar z_j}&\rightarrow&\frac{(-)\delta^\beta_\alpha}{-\beta(z_i-z_j)}=\frac{\delta^\beta_\alpha}{\beta(z_i-z_j)}
\end{eqnarray}
thus, we get the gauge limit \eqref{limit} used above:
\begin{eqnarray}
\frac{1}{z}&\rightarrow&\frac{1}{z}\nonumber\\
\frac{1}{\bar z}&\rightarrow&\frac{1}{\beta z}
\end{eqnarray}
(4) The zero mode of pure spinor formalism  for tree amplitudes is independent of HSZ gauge, namely,
\begin{eqnarray}
\langle(\lambda\gamma^m\theta)(\lambda\gamma^p\theta)(\lambda\gamma^q\theta)(\theta \gamma_{mpq}\theta)\rangle=1,
\end{eqnarray}
which holds for both chiral and anti-chiral part of the amplitude. It is already used in closed string calculations.

Now it is easy to check that the heterotic string amplitude is independent of chirality condition of the supersymmetric (SUSY) and bosonic part. By using rule (4), it is obvious to obtain
\begin{eqnarray}
\mathcal{A}_{H4}&=&\int d^2 z_4\, \mathcal{A}_{KN}(z_{ij})\mathcal{A}_J(\bar z_4)\mathcal{A}_O(z_4)\nonumber\\
&=&\int dz_4\, \frac{\beta}{\alpha'}\,\delta(\sum_{j=1}^3\frac{k_4\cdot k_j}{z_{4j}})\frac{1}{\beta}\mathcal{A}_J(z_4)\mathcal{A}_O(z_4)\nonumber\\
&=&\int d^2 z_4\, \mathcal{A}_{KN}(z_{ij})\mathcal{A}_J(z_4)\mathcal{A}_O(\bar z_4)
\end{eqnarray}
namely,
$${\bf Heterotic}\,\,\overline{\bf susy}\times {\bf bosonic}\Longleftrightarrow {\bf Heterotic\,\,susy}\times \overline{\bf bosonic}$$

\section{Conclusions}
In this paper, we showed the BRST equivalence between MS prescription \cite{GY} and ours, and we obtain the amplitude without explicit $\alpha'$ dependence rather than $\alpha'\rightarrow\infty$ or $\alpha'\rightarrow0$ limit. Instead of insertion of $\delta$ functions in vertex operator we introduce the formula with the dependence on both holomorphic and anti-holomorphic part just like the standard calculations of closed and heterotic strings in conformal gauge. Then, after taking the singular gauge limit, the integration of Koba-Nielsen factors give us the constraints in $\delta$-functions which coincides with the CHY scattering equations; meanwhile, by changing the boundary conditions, the rest of the anti-holomorphic part turns into the holomorphic part times polynomials of singular gauge parameters, which cancels the gauge dependence inside $\delta$-functions. 

Our method could be easily generalized to $N$-point scattering amplitude with $g$ loops by:\\
(a) Calculate the amplitude in usual conformal gauge;\\
(b) Take some singular gauge limit and then integrate over $\bar z$, which will only give us the $\delta$-functions;\\
(c) The integral over $z$ will lead to substitution of the solutions with respect to the constraints inside $\delta$-functions (scattering equations) into the integrand.\\
We could also propose that those singular gauge limits are related to some special quasi-conformal symmetry by Beltrami derivatives.

\section{Acknowledgments}

Y.L. would like to thank Yao Ma and Di Wang for helpful discussions with them especially at tea time at Simons Center for Geometry and Physics. W.S. was supported in part by National Science Foundation Grant No. PHY-1620628.

\end{document}